\documentclass[pra,twocolumn, showpacs, showkeys, secnumarabic, aps, amsmath, amssymb, nofootinbib, superscriptaddress, longbibliography, floatfix, table-of-contents, dblfloatfix]{revtex4-2}

\usepackage[utf8]{inputenc}
\usepackage[pdftex]{graphicx}
\usepackage{mathrsfs}
\usepackage[colorlinks, breaklinks, urlcolor={blue}, linkcolor={red}, citecolor={blue}]{hyperref}
\usepackage{array}
\usepackage{amsmath}
\usepackage{type1cm}
\usepackage{lettrine}
\usepackage[english]{babel}
\usepackage{lmodern}
\usepackage{microtype}
\usepackage{booktabs}
\usepackage[T1]{fontenc}
\usepackage[boxed, vlined]{algorithm2e}
\usepackage{caption}
\usepackage{braket}
\usepackage{xcolor}
\usepackage{orcidlink}

\begin{document}

\title{Entanglement preservation in tripartite quantum systems under dephasing dynamics }

\author{Chandrashekar Radhakrishnan \orcidlink{0000-0001-9721-1741}}
\email{chandrashekar.radhakrishnan@nyu.edu}
\affiliation{Department of Computer Science and Engineering, New York University Shanghai, 567
West Yangsi Road, Pudong, Shanghai 200124, China}

\author{Sovik Roy \orcidlink{0000-0003-4334-341X}}
\email{s.roy2.tmsl@ticollege.org}
\affiliation{Department of Mathematics, Techno Main Salt Lake (Engg. Colg.), Techno India Group, EM 4/1, Sector V, Salt Lake, Kolkata  700091, India}

\author{Ravikumar Chinnarasu \orcidlink{0000-0002-3912-4583}}
\email{chinnarasu@wisc.edu}
\affiliation{Department of Physics, University of Wisconsin-Madison, Madison, Wisconsin 53706, USA}

\author{Md.~Manirul Ali \orcidlink{0000-0002-5076-7619}}
\email{manirul@citchennai.net}
\affiliation{Centre for Quantum Science and Technology, Chennai Institute of Technology, Chennai 600069, India}

\begin{abstract}
Protecting entanglement from decoherence is a critical aspect of quantum information processsing.
For many-body quantum systems evolving under decoherence, estimating multipartite entanglement is challenging.
This challenge can be met up by considering distance based measure such as relative entropy of entanglement which
decisively measures entanglement in both pure as well as mixed states. In this work, we investigate the tripartite
entanglement dynamics of pure and mixed states in the presence of a structured dephasing environment at finite
temperature. We show that the robustness of the quantum system to decoherence is dependent on the distribution
of entanglement and its relation to different configurations of the bath. If the bath is structured individually such
that each qubit has its own environment, the system has different dynamics compared to when the bath is common
to all the three qubits. From the results we conjecture that there is a connection between the distribution of
entanglement among the qubits and the distribution of bath degrees of freedom, and the interplay of these two
distributions determines the decay rate of the entanglement dynamics. The sustainability of tripartite entanglement
is shown to be enhanced significantly in presence of reservoir memory.
\end{abstract}


\maketitle

\noindent
\section{Introduction}\label{sec:intro}

Entanglement is a nonclassical feature of multipartite quantum systems \cite{horo865}.  It has practical applications in information
processing tasks such as teleportation \cite{bennett1895}, superdense coding \cite{bennett2881}, cryptography \cite{gisin74} and
computing \cite{nielsen2010}. Thus entanglement is a resource and the success of quantum information protocols relies on our
ability to retain it. However, quantum systems interact with the surrounding environment and undergoes decoherence which in
turn results in the loss of quantum features like entanglement and coherence \cite{zurek715,breuer2002}. Several methods have
been developed to circumvent the adverse effect of decoherence and some of them are quantum error correction
\cite{steane1996,cory1998}, finding the decoherence free subspace \cite{zanardi1997,lidar1998}, and dynamical decoupling
\cite{viola1999,Lloyd1999}. Apart from these well known methods, there is a relatively new approach known as environment
engineering or reservoir engineering \cite{Braun2002,Sarlette2011,Nokkala2016,mazzola79,ankim2010}. In our work we adopt
this approach where we engineer the coupling between the system and the environment to enhance the longevity of the quantum
states. Using this approach we design the reservoirs such that the memory effects help in slowing down the decay process.  This helps us
to classify the multipartite entangled states based on their robustness to decoherence. For a tripartite pure state, the entanglement
can be calculated using measures like von-Neumann entropy and $3$-tangle. However for mixed state, von-Neumann entropy
is no longer useful. Although the $3$-tangle can be used, it needs to be optimized over all possible pure state decomposition
which is mathematically hard to perform. An open quantum system which is initially pure, becomes a mixed state during their
course of temporal evolution. Hence we need an entanglement measure which works uniformly for both pure and mixed states.
The only class of measures which satisfies this requirement is the distance based measure of entanglement. Of the distance based
entanglement measures, we use the well known relative entropy of entanglement.

The quantum dynamics of a system in contact with an environment is a non-unitary process. This dynamics can be either Markovian
or non-Markovian \cite{BreuerRMP16,deVega17}.  Generally the Markovian and non-Markovian dynamics of the two-qubit entangled
states have very distinctive features  \cite{mazzola79,bellomo99,wang78,dajka2008,paz100,junli82,ali82}. An investigation of the
entanglement dynamics of multipartite systems with more than two qubits is a topic of active research in recent years
\cite{ankim2010,ys2010,ma762007,aolita2008,lopez2008101,weinstein2009,christo2014}. One of the barriers in such an
investigation is to identify a good measure of entanglement which works equally for both pure and mixed states. In this work we
use a relative entropy of entanglement to investigate the multipartite entanglement dynamics of a tripartite system in the presence
of a structured dephasing environment. We consider the two possible scenarios {\it (a)} Each qubit interacts with its own environment
and {\it (b)} The qubits collectively interact with a common environment.  The sustainability of tripartite entanglement for different
pure and mixed states is examined under the following conditions:  {\it (i)} local Markov  {\it (ii)} local non-Markov
{\it (iii)} common Markov and {\it (iv)} common non-Markov dephasing environment.

\section{Overview on Multipartite Entanglement Measures}\label{sec:measures}

Entanglement in bipartite systems is measured using concurrence \cite{wootters801998}. When we are investigating the time dynamics tripartite
systems, though we consider the initial states to be pure states, they proceed to evolve into mixed states. Hence any entanglement measure we
use should be capable of computing the entanglement for both pure and mixed states with relative ease. In the case of multipartite systems
there were several attempts to establish entanglement measures similar to the bipartite concurrence.  Some of these measures are the tangle
[ for tangle], $n$-tangle \cite{coffman2000}, generalized concurrence [for generalized concurrence], negativity \cite{vidal652002,leechi2003}.
While the entanglement of tripartite pure states can be calculated using tangle, for mixed state its computation requires an optimization
over all pure state decompositions.  This optimization is a computationally complex procedure.  Similarly, the computation of generalized
concurrence requires solving a complex optimization problem.  Negativity is a necessary and sufficient condition in a bipartite system and
hence is widely used to evaluate entanglement for such systems.  Beyond bipartite systems, it is only a sufficient condition and hence is not
sufficient to characterize entanglement in tripartite systems.

The methods described above are all based on the intrinsic properties of entangled quantum states.  An alternative method is to consider
the Hilbert space of the quantum states and then compute the distance between the entangled state and their closest separable (nonentangled)
state.  The commonly used distance measure in this context is the relative entropy.  The relative entropy of entanglement measure which
quantifies how much a given entangled state can be distinguished from a separable state is defined as
\begin{eqnarray}
\label{REE1}
E(\rho) = \min_{\sigma \in \mathcal{D}}S(\rho \vert\vert \sigma),
\end{eqnarray}
where $\rho$ is an arbitrary density matrix for which we are trying to estimate the entanglement, $\sigma$ is a separable state and
$\mathcal{D}$ is the set of all separable states.  Here the quantum relative entropy is
\begin{eqnarray}
\label{REE2}
S(\rho \vert\vert \sigma) = {\rm Tr} \{ \rho(\ln \rho - \ln \sigma) \},
\end{eqnarray}
This measure $E(\rho)$ gives the amount of entanglement in a state $\rho$ as a distance to its closest separable state $\sigma$.
The more entangled a quantum state is, the more it is distinguishable and consequently is much farther away from a separable state.

\section{Description of the Models}\label{sec:Model}

\subsection{Three Qubits Under Local Dephasing Environment}\label{subsec:Local}
\begin{figure}[h]
\includegraphics[width=0.95 \columnwidth]{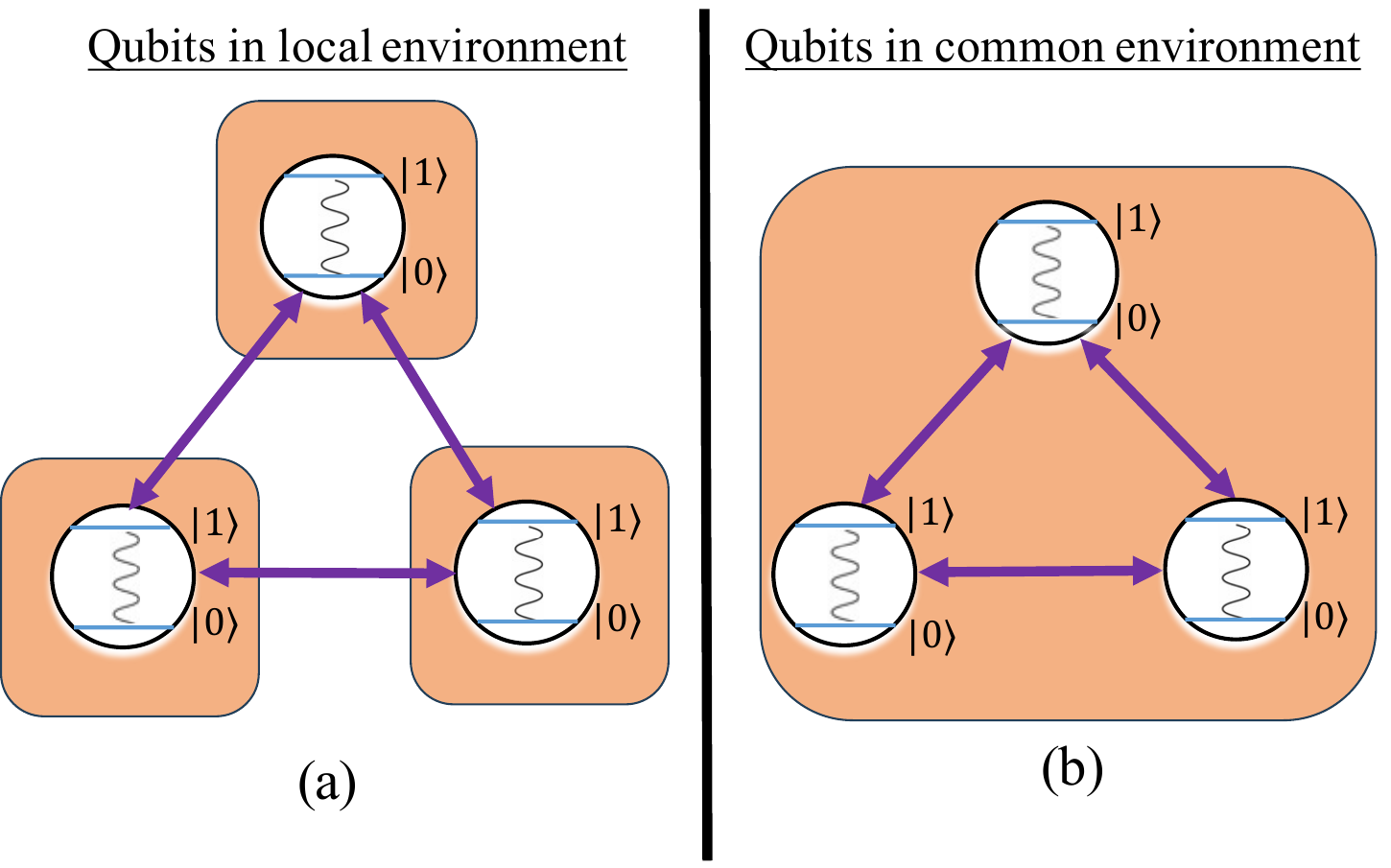}
\vskip -0.2cm
\caption{A schematic figure of a tripartite system in (a) local environment and (b) common environment.  The three qubits in the
system are entangled states. }
\label{fig0}
\end{figure}

We consider a spin-boson model of three non-interacting qubits, each of which is interacting with a local bosonic reservoir
as shown in Fig.~\ref{fig0}(a). The total Hamiltonian of the three qubits and the environment is given by
\begin{eqnarray}
\label{DephL}
H = \sum_{i=1}^{3} \bigg[ \frac{\hbar}{2}  \omega_0^i \sigma_z^i + \sum_k \hbar \omega_{ik} b_{ik}^{\dagger} b_{ik}
+ \sigma_z^i \left( B_i + B_i^{\dagger} \right) \bigg]
\end{eqnarray}
with $B_i = \sum_k g_{ik} b_{ik}$, where $\omega_0^i$ and $\sigma_z^i$ are respectively the transition frequency and the Pauli spin operator
associated with the $i^{th}$ qubit. For simplicity, we assume that all three qubits have the same transition frequencies $\omega_0^i=\omega_0$.
The local $i^{th}$ environment is modeled as a collection of infinite bosonic modes with frequencies $\omega_{ik}$.  Here $b_{ik}^{\dagger}$
($b_{ik}$) is the creation (annihilation) operator associated with the $k^{th}$ mode of the local environment interacting with the $i^{th}$
qubit. The factor $g_{ik}$ represents the coupling strength between the $i^{th}$ qubit and its local environment. In the continuum limit,
$\sum_k |g_{ik}|^2 \rightarrow \int d\omega J_{i}(\omega) \delta(\omega_{ik}-\omega)$, where $J_{i}(\omega)$ is the spectral density associated with the local environment of the $i^{th}$ qubit. Initially each of the three qubit system is decoupled from their respective
environments and each local environment `$i$' is in thermal equilibrium at a temperature $T_{i}$. The system and the environment can
be considered as a composite quantum object evolving under the Hamiltonian $H$. We need to trace out the environment degrees of
freedom and obtain the reduced density matrix corresponding to the quantum system to investigate its dynamics.  For the tripartite system,
the quantum master equation describing the decay dynamics under a local dephasing is
\begin{eqnarray}
\label{NL}
\frac{d}{dt} \rho(t) = \sum_{i=1}^{3} \gamma_{i}(t)  \Big( \sigma_z^i \rho(t) \sigma_z^i - \rho(t) \Big),
\end{eqnarray}
where $\rho(t)$ is the density matrix of the system and the time-dependent factor $\gamma_{i}(t)$ is given below:
\begin{eqnarray}
\label{GTD}
\gamma_{i}(t) = 2 \int_{0}^{\infty} d\omega  J_{i}(\omega) \coth\left(\frac{\hbar \omega}{2 k_B T_i}\right) \frac{\sin(\omega t)}{\omega}.
\end{eqnarray}
The time-dependent dephasing rate $\gamma_{i}(t)$ is determined by the spectral density $J_{i}(\omega)$. For the dephasing model we
consider the Ohmic-type spectral density \cite{leggett1987} given below:
\begin{eqnarray}
J_{i}(\omega) = \eta_{i} \omega \exp\left(-\frac{\omega}{\Lambda_i} \right).
\label{ohm}
\end{eqnarray}
Under normal circumstances the environment is large enough to quickly reset it back to its initial state.  In our work we examine the
reservoir memory effect on the entanglement dynamics of a tripartite system when the spectral density $J(\omega)$ has a finite cut-off
frequency $\Lambda_{i}$. In the Markov approximation, correlation time of the environment is much smaller than the time scale of the system
dynamics.  Under that condition one can replace the time dependent coefficient $\gamma_{i}(t)$ by its long-time Markov value
$\gamma_{i}^M=4 \pi \eta_i k_B T_i/\hbar$ and the dynamics of the system density matrix is referred to as Markovian dynamics.
If we assume uniform spectral densities with uniform coupling strengths $\eta_{i} = \eta$, and also all the individual
environments are at the same temperature $T_{i} = T$, then the decay constant $\gamma_{i}^M$ reduces to $\gamma_0=4 \pi \eta k_B T/\hbar$.
We note that the Markovian assumption is valid under the situation when the relaxation time of the bath is very short compared to the
evolution time of the system.  When the bath relaxation time and the system evolution time are comparable to each other, then the
reservoir memory is not negligible and the entanglement dynamics of three qubits is governed by the master equation Eq.~(\ref{NL}), where
the time dependent dephasing rate $\gamma(t)$ is given by Eq.~(\ref{GTD}).

\subsection{Three Qubits Under Common Dephasing Environment}\label{subsec:Common}

Next, we consider the situation where all the three qubits are under the interaction with a common reservoir as depicted
in Fig.~\ref{fig0}(b). The Hamiltonian for the tripartite qubit systems coupled to a common environment is given by
\begin{eqnarray}
\label{DephC}
H = \frac{\hbar}{2} \sum_{i=1}^{3} \omega_0^i \sigma_z^i + \sum_k \hbar \omega_{k} b_{k}^{\dagger} b_{k}
+ S_z \left( B + B^{\dagger} \right).
\end{eqnarray}
We assume that all three qubits have same transition frequency $\omega_0^i=\omega_0$.  The common environment is
modelled as a collection of bosonic field modes with frequencies $\omega_{k}$, where $b_{k}^{\dagger}$ ($b_{k}$) is
the creation (annihilation) operator associated with the $k$th mode of the environment. The reservoir operator
$B=\hbar \sum_k g_{k} b_{k}$, where $g_{k}$ is the coupling strength between the tripartite system and the common
environment. The operator $S_z = \sum_i \sigma_z^i$ is the collective spin operator of the three-qubit system.
The common environment is in thermal equilibrium at a temperature $T$ and initially the tripartite state and the
environment are decoupled such that the collective state of the system and environment can be expressed as a product
of their corresponding reduced density matrices. The quantum master equation of the tripartite state in contact with a
common dephasing environment at finite temperature is:
\begin{eqnarray}
\frac{d}{dt} \rho(t)\!=\!\gamma(t) S_z \rho(t) S_z\!-\!\alpha(t) S_z S_z \rho(t)\!-\!\alpha^{\ast}(t) \rho(t) S_z S_z,
\label{NC}
\end{eqnarray}
where
\begin{eqnarray}
\nonumber
\alpha(t) &=& \int_{0}^{\infty} d\omega  J(\omega) \coth\left(\frac{\hbar \omega}{2 k_B T}\right) \frac{\sin(\omega t)}{\omega} \\
&-& i \int_{0}^{\infty} d\omega  J(\omega) \frac{1-\cos(\omega t)}{\omega},
\end{eqnarray}
and $J(\omega)=\eta \omega \exp (-\omega/\Lambda)$.  When the reservoir memory is negligible, Eq.~(\ref{NC}) can be reduced using
the Markov approximation and the corresponding quantum master equation takes the form:
\begin{eqnarray}
\frac{d}{dt} \rho(t) = \frac{\gamma_0}{2}  \Big( 2 S_z \rho(t) S_z  - S_z S_z \rho(t)
-  \rho(t) S_z S_z \Big).
\label{MC}
\end{eqnarray}
The non-Markovian dynamics Eq.~(\ref{NC}) in presence of reservoir memory is significantly different from its memoryless Markovian
counterpart Eq.~(\ref{MC}) which is valid only under a situation when the reservoir correlation functions decay very fast.

Considering specific three qubit state, we show how reservoir memory effects can improve the robustness of the multiqubit entanglement
under common environment. The entanglement dynamics is investigated for both Markov and non-Markov dephasing dynamics.
In our numerical calculations we choose $k_B T=\hbar \omega_0/4\pi$ for which $\gamma_0=\eta \omega_0$.
Also we consider the following parameter values of $\eta=0.1$, and $\Lambda=10^{-2}\omega_0$.
\begin{figure}[h]
\includegraphics[width=\columnwidth]{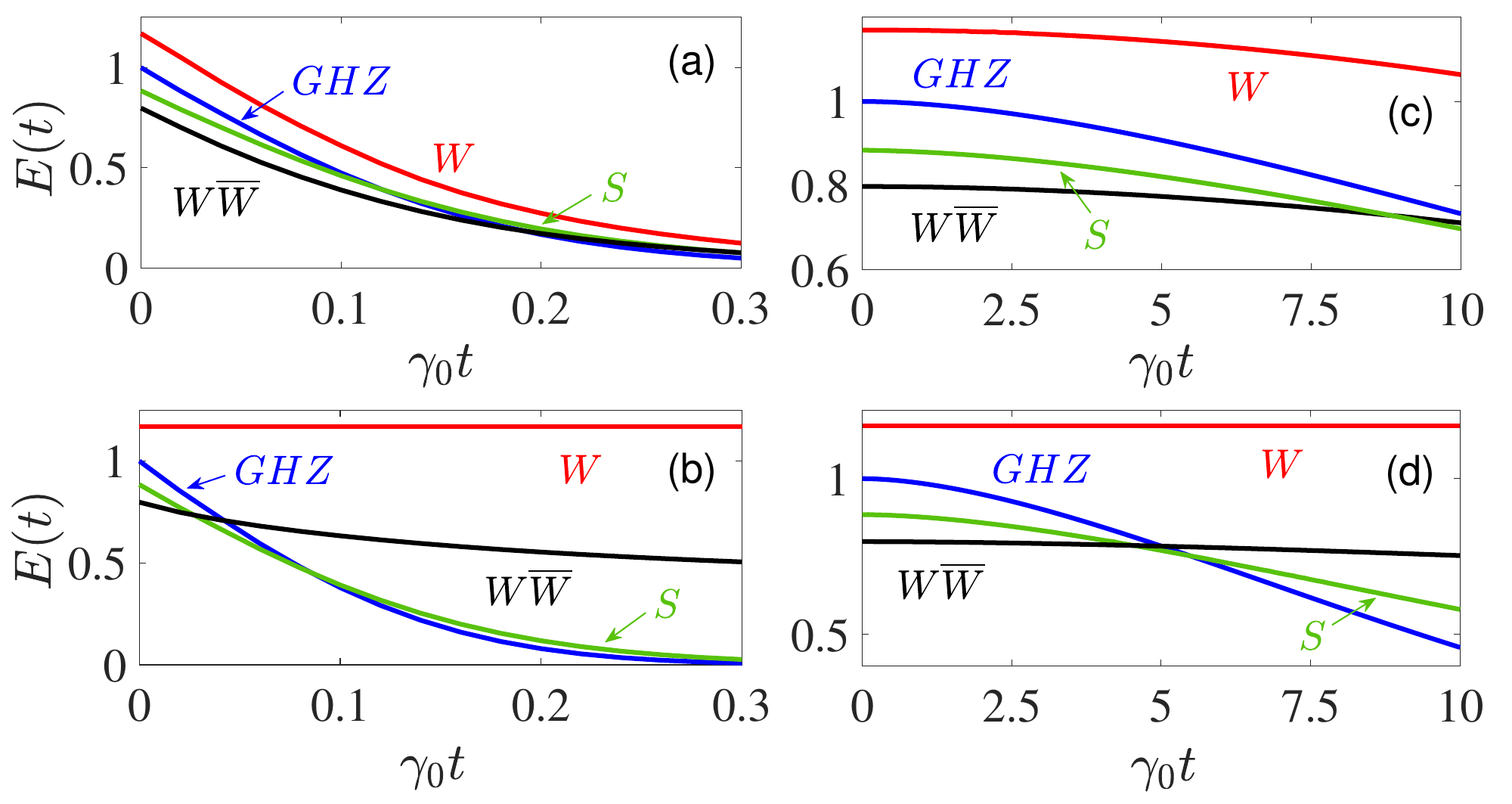}
\vskip -0.2cm
\caption{The entanglement dynamics of the GHZ, W, $W\overline{W}$ and Star states in a dephasing environment is given
for (a) local Markov, (b) common Markov, (c) local non-Markov, and (d) common non-Markov environments.}
\label{fig1}
\end{figure}
\begin{figure}[h]
\includegraphics[width=\columnwidth]{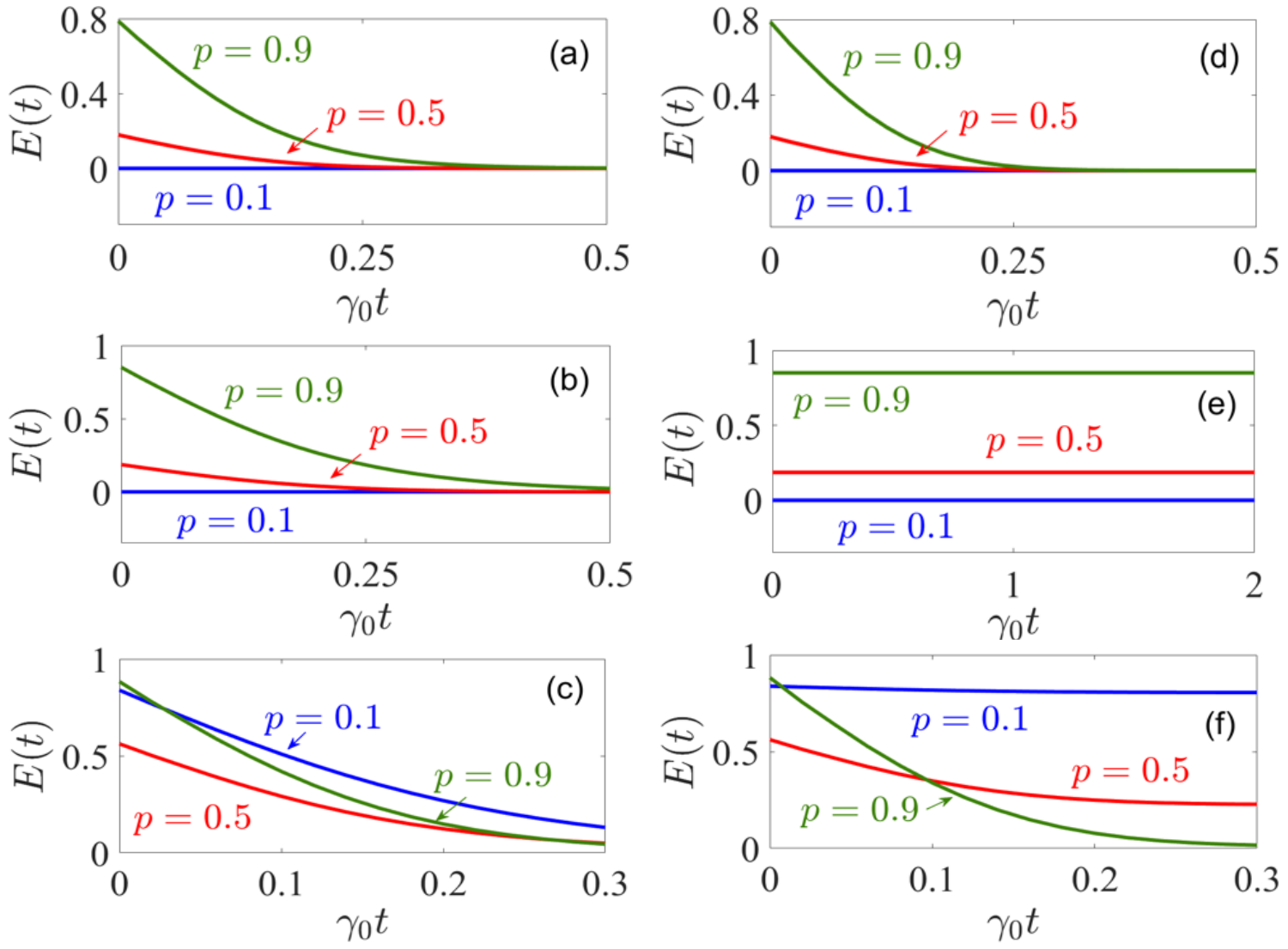}
\vskip -0.2cm
\caption{The Entanglement dynamics plot of tripartite mixed states under Markovian dephasing environment is given.
For (a) Werner-GHZ (b) Werner-W and (c) mixture of GHZ and W-state, we show the plot of local Markov environment.
The common Markov environment plots are given for (d) Werner-GHZ (e) Werner-W and (f) mixture of GHZ and W-state.}
\label{fig2}
\end{figure}
\begin{figure}[h]
\includegraphics[width=\columnwidth]{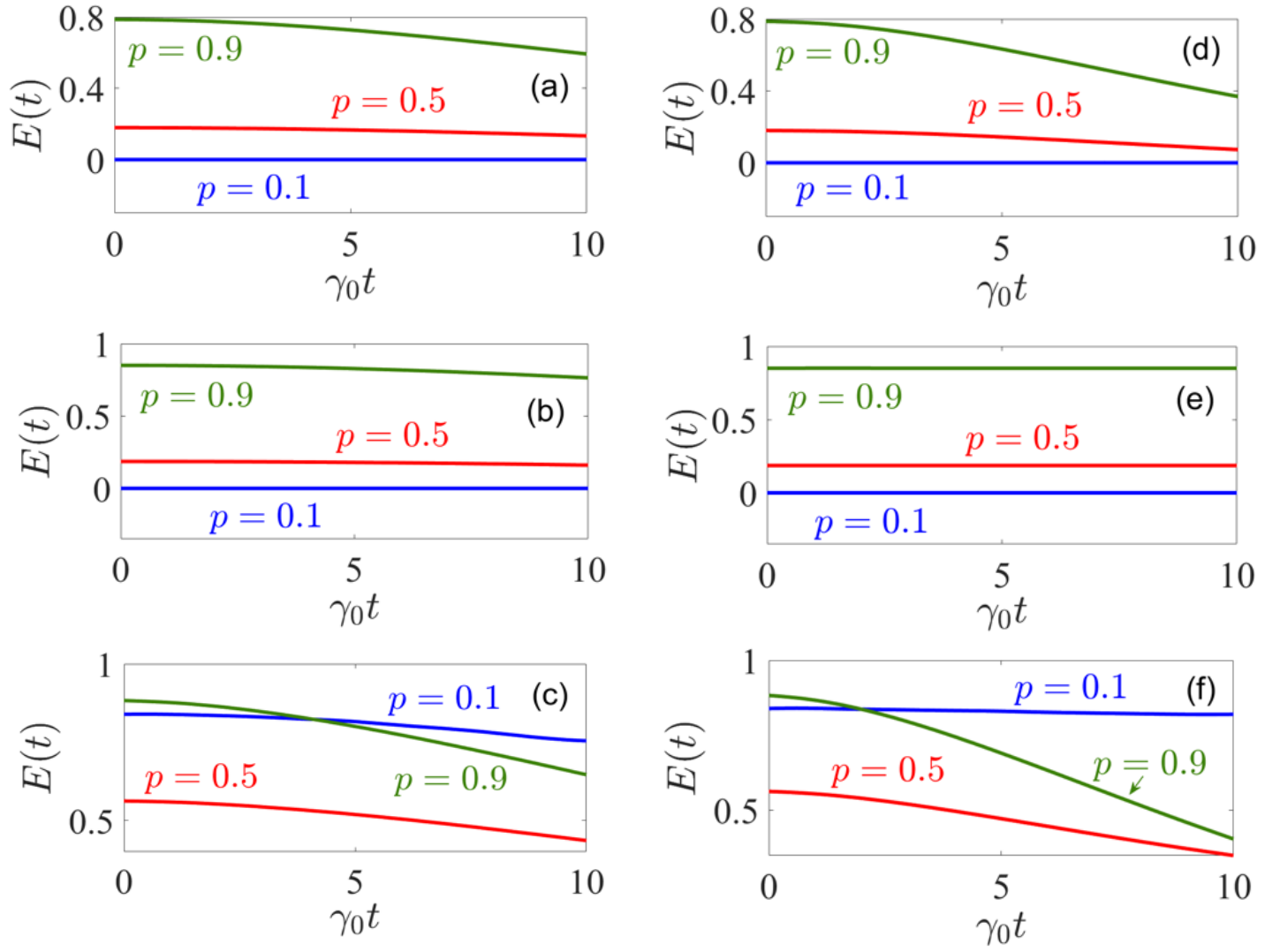}
\vskip -0.2cm
\caption{The entanglement dynamics of tripartite mixed states under non-Markovian dephasing environment is shown
for local environment for the mixed states {\it viz } (a) Werner-GHZ (b) Werner-W and (c)  mixture of GHZ and W-state
and for the common environment again for (d) Werner-GHZ (e) Werner-W and (f) mixture of GHZ and W-state.}
\label{fig3}
\end{figure}

\subsection{Quantum states}\label{subsec:quantumstates}
The entanglement dynamics of specific three qubit systems like GHZ, W, $W\overline{W}$ and Star states under local and common dephasing
environment is studied using the relative entropy of entanglement.   Based on SLOCC (Stochastic Local Operations and Classical Communication)
operations the tripartite states are classified into two classes {\it viz}, the GHZ class and the W class \cite{Durrcirac}.
The tripartite GHZ and W states
\begin{eqnarray}
\label{ghzstate}
\vert GHZ\rangle &=& \frac{1}{\sqrt{2}}\Big(\vert 000\rangle + \vert 111\rangle\Big), \\
\label{wstate}
\vert W\rangle &=& \frac{1}{\sqrt{3}}\Big(\vert 001\rangle + \vert 010\rangle + \vert 100\rangle \Big).
\end{eqnarray}
are quite different in their entanglement distribution. The three qubit GHZ state consists of purely tripartite entanglement which vanishes
completely when even one of the qubit undergoes decoherence.  In the case of the W state the entanglement is distributed completely in a bipartite
way. There is no tripartite entanglement in the W state and this can be concluded from the vanishing of $3$-tangle which is a quantifier of
tripartite entanglement.

Since the GHZ and W states have mutually exclusive configuration in terms of distribution of entanglement, we investigate the $W \overline{W}$
state which has both tripartite and bipartite distribution of entanglement.  The form of the tripartite $W \overline{W}$ is
\begin{eqnarray}
\label{wwbar1}
\vert W\overline{W}\rangle = \frac{1}{\sqrt{2}}(\vert W\rangle + \vert \overline{W}\rangle).
\end{eqnarray}
where $\vert \overline{W}\rangle = \frac{1}{\sqrt{3}}(\vert 110\rangle + \vert 101\rangle + \vert 011\rangle)$ is the spin flipped W state
\cite{chandra2020}.  The $3$-tangle of the $W \overline{W}$ state is equal to $1/3$ and the entanglement of the reduced bipartite state as
measured using concurrence is $C=1/3$ and this proves that this state has both bipartite and tripartite distribution of entanglement.

The states considered above namely the GHZ state, the W state and the $W \overline{W}$ are all symmetric states and their reduced
bipartite entanglement does not depend on which qubit has undergone decoherence.  Contrasting to these types of states is the Star state
\cite{buzek67,anders2006}
\begin{equation}
\label{star1}
\vert S \rangle = \frac{1}{2}[\vert 000\rangle + \vert 100\rangle + \vert 101\rangle + \vert 111\rangle].
\end{equation}
in which there is a central qubit which is entangled to the other two peripheral qubits and the two peripheral qubits are not entangled with each
other in a bipartite manner.  The tripartite entanglement of the Star state computed using the $3$-tangle is $1/4$ and the different bipartite
entanglements measured using concurrence are $C_{AC} = C_{BC} = 1/2$ and $C_{AB} = 0$ and this illustrates the asymmetric star type distribution
of entanglement.  Thus the $W \overline{W}$ and Star states are good test beds to study the entanglement dynamics when the entanglement itself
is distributed at different levels in the system.

\section{Entanglement dynamics of pure states}
The entanglement dynamics of tripartite pure states in local and common environments is discussed below for the dephasing model.

\subsection{Markovian dephasing environment}
The Markovian dynamics of the GHZ, W, Star and $W\overline{W}$ states is shown in Fig.~\ref{fig1} for both the local and common environments. The dynamics of entanglement measured using the relative entropy of entanglement under Markovian local dephasing environment is given in Fig.~\ref{fig1}(a). The amount of entanglement of the initial states (at $t=0$) can be given through the ordering relation
$E(W) > E (GHZ) > E (Star) > E (W\overline{W})$.  The entanglement decays exponentially with time and vanishes for all these states
at $\gamma_{0}t = 0.5$.  We observe that for all the tripartite systems when each qubit is in contact with a localized individual bath
the entanglement decay is exponential throughout indicating that the decay is uniform for both bipartite and tripartite distribution.

The time variation of entanglement of the tripartite pure states in a common Markov dephasing environment is given in Fig.~\ref{fig1}(b).
The entanglement dynamics of the GHZ and star states are exponential and it is very similar to the dynamics in a local Markovian bath. The
entanglement of these two states in the common bath fall faster than that in the local Markov environment.  Beyond $\gamma_{0}t = 0.3$, the
entanglement of these states vanish.  In the case of $W\overline{W}$ state the entanglement decay is not exponential and is very slow compared
with the GHZ and the Star states.  This behavior is in contrast with the results obtained for the decay dynamics of the Makov local bath.
The most interesting behavior is being exhibited by the W states, for which we find that the entanglement remains constant over time.
It is well known that the entanglement in W states is distributed only in a bipartite fashion, whereas in the remaining states, the
entanglement is distributed completely in a tripartite manner for the GHZ state and in both tripartite and bipartite manner for the W states
and the Star states. These observations lead us to believe the bipartite entanglement is less affected by the common environment than the
tripartite entanglement.

\subsection{Non-Markovian dephasing environment}
There are two important time scales in an open quantum system and they are (a) the evolution time of the system $\tau_{s}$ and (b) the
correlation time of the bath $\tau_{b}$.  When $\tau_{s} \gg \tau_{b}$, the dynamics is Markovian and the evolution of the states under
this condition is discussed in the previous subsection.  On the contrary when the two time scales are comparable {\it i.e.,}
$\tau_{s} \approx \tau_{b}$, the system exhibits non-Markovian dynamics.  The evolution of the pure states in the non-Markovian
local dynamics is given in Fig.~\ref{fig1}(c). We find that the entanglement decays in all the quantum states under investigation.
But the rate of decay is much more slower than that for a local Markov process which has been discussed in the previous section
(see Fig.~\ref{fig1}(a) for comparison). This is because in a non-Markov process, there is a backflow of information from the
environment to the system which essentially slows down the decoherence. In terms of the tripartite states, the W-states and the
$W\overline{W}$ states experience slower decay than the GHZ and the star states. Overall we conclude that the entanglement
decay of a non-Markov process is slower than that of a Markov process.

The non-Markovian dynamics of the tripartite pure states in contact with a common dephasing environment is given in Fig.~\ref{fig1}(d).
We find that entanglement of GHZ and Star states decreases with $\gamma_{0} t$.  Out of these two states, the decay rate of GHZ state
is much higher than that of the star state.  From a comparison with the dynamics of local non-Markov environment in Fig.~\ref{fig1}(c)
we find that the entanglement of GHZ states and Star states fall faster when they are in a common environment.  On the contrary, the
entanglement of $W$ and $W\overline{W}$ states remains constant in time $(\gamma_{0} t)$ when they are exposed to a common
environment. But from Fig.~\ref{fig1} (c) we observe that when the states are interacting with a local environment the
$W$ and $W\overline{W}$ states exhibits a decay.  So we find that in a tripartite system interacting with a common dephasing bath,
when there is a symmetric distribution of bipartite entanglement, there is no entanglement decay.

\section{Mixed states under dephasing environment}
In so far, the investigations have been carried out only for the pure states.  Under general circumstances, it is hard to prepare
and sustain pure states.  Hence for the sake of completeness we need to study the entanglement dynamics in  mixed states as well.
Towards this end we consider the following mixed states namely {\it (i)} Werner-GHZ state, {\it (ii)} Werner-W state and
{\it (iii)} mixture of GHZ and W states \cite{junge2009,lohmayer97}
\begin{eqnarray}
    \rho_{wr}(GHZ,p) &=& p\vert GHZ\rangle\langle GHZ\vert + \frac{1-p}{8}\mathbb{I}_{8},
    \label{WernerGHZ} \\
    \rho_{wr}(W,p) &=& p\vert W\rangle\langle W\vert + \frac{1-p}{8}\mathbb{I}_{8},
    \label{WernerW}
\end{eqnarray}
\begin{equation}
   \rho_{_{GW}} = p\vert GHZ\rangle\langle GHZ\vert + (1-p)\vert W\rangle\langle W\vert.
   \label{GHZWerner}
\end{equation}
Here $p$ is the probability of mixing and $\mathbb{I}_{8}$ is the $8 \times 8$ identity matrix.  Of these, the first two states are
regular tripartite states mixed with white noise (maximally mixed state). The third state Eq.~(\ref{GHZWerner}) is a statistical mixture
of $GHZ$ and $W$ states.  In this state by changing the value of $p$, we can vary the amount of bipartite and tripartite entanglement
in the system.  The entanglement dynamics of these states is examined when they are in contact with a local environment as well as
with a common environment.

\subsection{Markovian dephasing environment}
The variation of entanglement with time is given in Fig.~\ref{fig2}(a)-(c) for the Werner-GHZ, Werner-W and the GHZ-W
mixed state respectively, when these states are in contact with a local bath. From the plots we find that for the Werner-GHZ and the
Werner-W states the entanglement falls exponentially with $\gamma_{0} t$, for values of $p=0.9$ and $p=0.5$. In the case of
$p=0.1$, there is no entanglement in the initial state and so the entanglement remains zero throughout the evolution. For the state
which is a mixture of GHZ and W, the entanglement falls exponentially for all values of $p$. But lower the value of $p$, the
slower the fall in entanglement. This is because for the low values of $p$, the mixture consists of higher proportion of W-states.
As shown in Fig.~\ref{fig1}(a), the W-states with bipartite entanglement decays slower than the tripartite entangled GHZ states.
Hence the mixture with higher proportion of W-state decays slower.

The entanglement dynamics for the three mixed states when they are in contact with a common bath is given in Fig.~\ref{fig2}(d)-(f).
From figure \ref{fig2}(d) we notice that the entanglement of the Werner-GHZ state falls exponentially for $p=0.9$ and $p=0.5$ with
a little faster speed for a common bath in comparison with the local environment. The entanglement of the $W$ state remains almost
constant (see Fig.~\ref{fig2}(e)) for all values of $p$. In the case of the GHZ-W mixed states as shown in Fig.~\ref{fig2}(f), the entanglement
falls exponentially for $p=0.9$, when the GHZ component is higher and a bit slower for $p=0.5$ when there is a equal mixture of GHZ and
W states. If the mixed state contains a higher proportion of W state like when the value of mixing parameter is $p=0.1$
the entanglement decay is so small that it appears to be almost constant with $\gamma_{0} t$.

\subsection{Non-Markovian dephasing environment}
The entanglement decay plot of the mixed states is given in Fig.~\ref{fig3}(a)-(c) for a local non-Markov dephasing environment.
For this model we find that the entanglement of the Werner-GHZ and Werner-W states falls with time and the fall is much
slower than the exponential fall observed for the Markov decay. The decay of the GHZ-W mixed state also falls slowly
with time and the rate of fall depends on the value of the mixing parameter $(p)$, where lower the value of $p$, slower is the fall.
In figures \ref{fig3}(d)-(f) we plot the entanglement dynamics for the common non-Markov dephasing evironment.
The Werner-GHZ state exhibits a decrease in entanglement with time and also with decrease in $p$ the mixing parameter.
In the Werner-W state the entanglement remains constant throughout the evolution and the system can be considered as a
decoherence free system for all values of the mixing parameter. In the GHZ-W mixed state, for low values of mixing parameter at
$p=0.1$ the entanglement is decoherence free. For higher values of the mixing parameter i.e., $p=0.5$ and $p=0.9$ the
entanglement exhibits an almost linear decay.

From the analysis of the Markov and non-Markov dynamics of the tripartite pure and mixed states in local as well in common environments,
we come to the conclusion that the W-states are decoherence free when they are in contact with a common environment. This can be observed
from the dynamics of the W-states. The amount of initial entanglement in a Werner-W state depends on the mixing probability $p$. This
is preserved when the states are in contact with a non-Markovian common bath.  Another validation comes from the analysis of the
GHZ-W mixture where we observe that if the mixture contains higher proportion of W-states the decay of the state is slower and
under the conditions when the state is maximally a W-state $(90 \%)$ the decoherence is negligible. This leads to the conclusion
that the W-states are decoherence free in a common dephasing environment.

\section{Conclusion}

Entanglement is an important resource of both theoretical and practical value in quantum information processing.  Any quantum
information protocol happens over a finite amount of time and due to this reason the dynamical stability of entanglement becomes an
important issue. Since quantum systems are affected by their environment, they decohere and consequently they loose their
entanglement. Relative entropy of entanglement has been considered here to study the dynamics of tripartite entangled states.
This measure can be used for both pure and mixed tripartite states without any problem.
Now an important question in this connection is how do we preserve the entanglement in a given system?  We can do so
by a careful choice of initial states and also by choosing to work under specific environmental conditions where the changes due to
the dynamics is negligible.  In this work we consider a tripartite system in an external dephasing environment.  Such dephasing
environment which is due to phase damping processes describe decoherence in realistic systems with practical implications.
Towards a comprehensive examination we consider four different pure states as well as three different mixed states and examine
their entanglement dynamics. Also we consider the two different situations where each qubit is in an environment of its own and
in a different situation when all the qubits are in a common environment. The entanglement dynamics depends on the
characteristic of the environment {\it viz.}, Markovian or non-Markovian. The non-Markovian entanglement dynamics
in the presence of reservoir memory is found to be significantly different from memoryless Markovian dynamics.

From our investigations we observe that the three pure states {\it viz} GHZ, Star and $W\overline{W}$ states exhibit decay of entanglement
with time both in Markovian and non-Markovian environment. The W-states decay in the memoryless Markov limit for both local
and common environment. In presence of reservoir memory the entanglement decays only for the situation where the three qubits have
their individual local baths. But if the qubits have a common bath the entanglement does not decay under non-Markovian dynamics.
Thus we can conclude that the W-states do not decohere in a non-Markovian common bath. To reinforce this idea we look at the dynamics of
Werner-W states as well as the GHZ-W mixed states.  The results show that for the Werner-W states, the entanglement initially present
remains constant throughout the evolution for non-Markovian dynamics.  The non-Markov dynamics of the GHZ-W mixture states show that
higher the amount of W states in this mixture slower is the decay.  From these results we have the following observations {\it viz}
{\it (a)} The W-states and $W\overline{W}$ states do not undergo decoherence when they are in common environment
{\it (b)} The rate of decay of GHZ and star staes is lower in the local non-Markov environment. This leads us to conjecture that there
is a connection between the distribution of entanglement among the qubits and the distribution of bath degrees of freedom (as either
local available only to selected qubit or global where it is available to all the qubits) and the interplay of these two distributions
determines the decay rate of the entanglement dynamics.

\begin{acknowledgments}
MMA was supported by the Centre for Quantum Science and Technology, Chennai Institute of Technology, India, via
funding number CIT/CQST/2021/RD-007.
\end{acknowledgments}


\bibliography{Entanglement3Qubit.bib}

\end{document}